\documentstyle[12pt]{article}
\newcommand{\be}{\begin{equation}}
\newcommand{\ee}{\end{equation}}
\newcommand{\bea}{\begin{eqnarray}}
\newcommand{\eea}{\end{eqnarray}}

\title{\Large \bf THE NON-PERTURBATIVE GROUNDSTATE OF Q.C.D. AND
THE LOCAL COMPOSITE OPERATOR $A^2_{\mu}$}
\author{\normalsize H. Verschelde, K. Knecht, K. Van Acoleyen
and M. Vanderkelen \\
\it University of Gent\\
\it Department of Mathematical Physics and Astronomy \\
\it Krijgslaan 281-S9 \\
\it B-9000 GENT, BELGIUM}
\date { }
\begin{document}
\maketitle
\pagenumbering{arabic}
\vskip 36pt
\begin{abstract}
We investigate the possibility that the dimension 2 condensate
$A^2_{\mu}$ has a non zero non-perturbative value in
Yang-Mills theory. We introduce a multiplicatively
renormalisable effective potential for this condensate and show
through two loop calculations that a non zero condensate is
energetically favoured.
\end{abstract}
\newpage

Recently [1] a lot of interest has arisen concerning the
possibility of a condensate in Yang-Mills-theory of mass
dimension 2. A possible candidate is given by the gauge
invariant operator
\bea
\Delta & = & \frac{1}{2} \frac{\langle 
\begin{array}[t]{c}
\min\\ \stackrel{\{U\}}{ } \end{array}
\int d^4 x (A^U_{\mu})^2 \rangle} {V.T.}\\ \nonumber
\mbox{} & = & \frac{1}{2} \langle \tilde{A}^2_{\mu} \rangle
\eea
where $\tilde{A}_{\mu} = A_{\mu}$ in the absolute Landau gauge [
11] and $A^2_{\mu} = A^a_{\mu}A^a_{\mu}$. There are several phenomenological reasons [2] to 
believe
that the groundstate of Q.C.D. favours a non-perturbative value
for this condensate different from zero. Theoretically, it was
shown in [1] that monopole condensation in compact Q.E.D. is
related to a phase transition for this condensate. In this paper
we would like to give further theoretical evidence that the
non-perturbative groundstate of Q.C.D. favourizes energetically
a non-zero value for this condensate. For this, several problems
have to be solved.

First of all, there is the question of what we mean by the
non-perturbative value of $\langle A^2_{\mu}\rangle$.
Perturbatively, this condensate is quadratically divergent and
Borel-non-summable because of the presence of ultraviolet
renormalons. If by non-perturbative, we mean that part of the
condensate that is proportional to the $\Lambda^2$-parameter,
this is ambiguous because it depends on an arbitrary summation
prescription for the perturbative part [3,4]. A second problem
is how to define a renormalisable effective potential for the
local composite operator $A^2_{\mu}$. Because the composite
operator is local, new divergences are introduced which
necessitate new counterterms that spoil an energy interpretation
[5]. In this letter, we will show how a unique non-perturbative
value of the condensate $\langle A^2_{\mu}\rangle$ can be
defined. For this condensate, we will construct a
multiplicatively renormalisable effective potential which is
unique and whose absolute minimum gives the non-perturbative
groundstate. We will calculate this effective potential up to
two loops and show that up to this order, the groundstate
favours a non zero value for the non-perturbative condensate
$\langle A^2_{\mu}\rangle$. We conclude with some numerical
results for the gluon condensate $\frac{\alpha_s}{\pi} \langle
F^2_{\mu \nu}\rangle$ and some comments.

To define the effective potential for the non-perturbative
condensate $\langle A^2_{\mu}\rangle$ we introduce a massterm
$\frac{1}{2} J(A^2_{\mu})$ in the SU(N) Yang-Mills Lagrangian in the Landau
gauge. This term generates new divergences in the generating
functional for connected Greens-functions W(J). There is a
quadratic divergence linear in J corresponding to the quadratic
divergence of $\langle A^2_{\mu}\rangle$. As we will show,
this divergence drops out of the effective potential so we don't
have to renormalise it. There is a logarithmic divergence linear
in J corresponding with multiplicative massrenormalization which
can be cancelled by a counterterm $\frac{1}{2}\delta Z_2 J A^2_{\mu}$.
Finally there is a logarithmic divergence quadratic in J which
corresponds to a new divergence in the Greens function $\langle
A^2_{\mu}(x) A^2_{\mu}(y)\rangle_c$ when $x \rightarrow
y$ and which can be cancelled by a counterterm $\delta \zeta
J^2/2$. These counterterms are sufficient to ensure a finite
renormalised W(J). The reader might question this on the basis
of the common wisdom that massive Yang-Mills-theory is
non-renormalizable [6]. However, the mass term $\frac{1}{2}J A^2_{\mu}$
is added to the Lagrangian after gauge fixing. Therefore, our
massive Lagrangian is not the one of massive Yang-Mills theory.
In particular, the vanDam-Veltman-Zakharov [7] discontinuity
theorem is not valid and we have a smooth $J \rightarrow 0$
limit. A simple power counting argument can then be used to show
that our new counterterms renormalise the theory. We will
discuss the problem of unitarity at the end of this letter.

Let us now try to define a non-perturbative value of $\langle 
A^2_{\mu}\rangle$. Therefore we consider the massive
gluonpropagator $G(k^2,J)$ as a function of J. Suppose
furthermore that G is a multivalued function of J. This means
that if one starts from the perturbative groundstate at J = 0
caracterised by a certain value of $\langle 
A^2_{\mu}\rangle$, makes a contour in the complex J-plane around
one or more singularities and then comes back to J = 0 on a
different Riemann sheet, one can end up in a non-perturbative
groundstate caracterised by a different value of the condensate.
This situation is analogous to $\lambda \phi^4$ theory with
negative mass term where $\langle \phi\rangle(J)$ is
multivalued. The role of the negative mass is played by the
tachyon pole, generated by infrared renormalons in the $
A^2_{\mu}$ channel. What is different is that in our case, there
is no spontaneous symmetry breaking and that the perturbation
series around the different vacua are identical. Then how can we
make a distinction between the perturbative and a
non-perturbative groundstate ? For that, we need a quantity
which is zero to all orders in perturbation theory. As a
candidate, we can take $G^{-1}(0,0)$ which because of gauge
invariance, is zero to all orders in perturbation theory. Hence
we can define the perturbative gluon propagator as the
propagator for which $\begin{array}[t]{c} \lim \\ \stackrel{J
\rightarrow 0}{ } \end{array} G^{-1}(0,J) = 0$. On the
perturbative Riemann sheet we have $G(k^2,J) = G_p(k^2,J)$ and
the perturbative condensate is then defined through :
\bea
\frac{1}{2} \langle A^2_{\mu}\rangle & = & \frac{1}{2} \int \frac{d^4
k}{(2\pi)^4} G_p(k^2,0) + \frac{1}{2} \int \frac{d^4
k}{(2\pi)^4} [G(k^2,J) - G_p(k^2,0)] \\ \nonumber
& = & \Delta_p + \Delta_{np}(J)
\eea
The perturbative part of the condensate $\Delta_p$ as defined is
this way, is not the perturbative series for $\frac{1}{2}\langle
A^2_{\mu}\rangle$ summed in some arbitrary way but the value of
$\frac{1}{2}\langle A^2_{\mu}\rangle$ for $J = 0$ on the perturbative
sheet. This perturbative value is well defined after
regularization and contains all the quadratic divergences. The
non-perturbative condensate is only logarithmically divergent
and vanishes with J on the perturbative sheet.

To construct an effective action for the non-perturbative
condensate $\Delta_{np}$,we consider the generating functional
W(J) and do a Legendre transform with respect to J. The only way
that W(J) can implicitly depend on $\Delta_p$ is through the
linear term in J which contains the quadratic divergences.
However, in the Legendre transform the linear terms in J cancel
so the Legendre transform of W(J) is implicitely only a function
of $\Delta_{np}$. Gauge invariance plays a very important role
in this. Indeed, because of gauge invariance, quadratic
divergences cancel in self-energy subdiagrams of W(J). So the
only possible dependence of W(J) on $\Delta_p$ is through the
overall quadratic divergence linear in J. As a consequence we
can forget about the perturbative condensate and use a gauge
invariant regularization such as dimensional regularization
where $\Delta_p$ is automatically zero. Introducing counterterms
$\frac{1}{2} \delta Z_2 J A^2_{\mu}$ and $\delta \zeta J^2/2$ for the
logarithmic divergences linear (multiplicative mass
renormalization) and quadratic in J (vacuum energy divergences)
we obtain a finite renormalised functional W(J) given by :
\be
e^{-W(J)} = \int [dA_{\mu}]\exp - \int d^D x \left[ \frac{1}{4}
F^2_{\mu \nu} + \frac{1}{2} Z_2 J A^2_{\mu} - (\zeta + \delta \zeta)
\frac{J^2}{2} + {\cal L}_{g.f} + {\cal L}_{c.t}\right]
\ee
To ensure a homogeneous renormalization group equation we had to
introduce a new independent parameter $\zeta(\mu)$. Defining the
bare quantities
\newpage
\bea
A^0_{\mu} & = & Z^{1/2}_3 A_{\mu} \nonumber \\
J_0 & = & \frac{Z_2}{Z_3} J \nonumber \\
g^2_0 & = & \mu^{\epsilon} \frac{Z_g}{Z^2_3} g^2 \nonumber\\
\zeta_0 J^2_0 & = & \mu^{-\epsilon}(\zeta + \delta \zeta)J^2
\eea
the RGE for W(J) becomes
\be
\left( \mu \frac{\partial}{\partial\mu} + \beta(g^2)\frac{\partial}{\partial
g^2} - \gamma_2(g^2)\int d^4 x J \frac{\delta}{\delta J}
+ \eta(g^2,\zeta)\frac{\partial}{\partial \zeta}\right) W = 0
\ee
where
\bea
\beta(g^2) & = & \mu \frac{\partial}{\partial \mu}
g^2|_{g_0,\epsilon} \nonumber \\
\gamma_2(g^2) & = & \mu \frac{\partial}{\partial \mu} \ln
\frac{Z_2}{Z_3}|_{g_0,\epsilon} \nonumber \\
\eta(g^2,\zeta) & = & \mu \frac{\partial}{\partial \mu}
\zeta|_{g_0,\epsilon, \zeta_0,J_0}
\eea
Because of (4) and the single valued relation
between $\mu$ and $g^2(\mu)$, we can consider $\zeta$ as a
function of $g^2$ and we have :
\bea
\mu \frac{\partial}{\partial \mu}
\zeta|_{g_0,\epsilon,J_0,\zeta_0} & = & \eta(g^2,\zeta)
\nonumber \\
& = & 2 \gamma_2 (g^2)\zeta + \delta(g^2)
\eea
where
\be
\delta(g^2) = (\epsilon + 2\gamma_2(g^2) - \beta(g^2)
\frac{\partial}{\partial g^2})\delta \zeta
\ee
is a finite function of $g^2$.

In defining a finite value for the energy functional W(J) we
have introduced two problems. First, since we had to introduce a
new parameter $\zeta$, there is a problem of uniqueness.
Secondly, for renormalisation purposes, we had to introduce a
quadratic term in J in the Lagrangian. Naively, one expects that
this will ruin an energy interpretation for the effective
potential defined via the Legendre transform.
In the case of the Gross-Neveu model [8], both problems were
solved by one of us in [9]. Concerning the first problem, it is
possible to choose $\zeta$ to be a unique meromorphic function
of $g^2$ such that if $g^2$ runs, $\zeta$ will run according to
(7). Indeed, the general solution of (7) reads
\be
\zeta(g^2) = \zeta_p(g^2) + \alpha \exp \left( 2 \int^{g^2}_1
\frac{\gamma_2(z)} {\beta(z)} dz \right)
\ee
$\zeta_p (g^2)$ is the particular solution of
\be
\beta(g^2) \frac{d}{dg^2} \zeta(g^2) = 2\gamma_2 (g^2) +
\delta(g^2) 
\ee
which has a Laurent expansion around $g^2 = 0$ :
\be
\zeta_p (g^2) = \frac{c_{-1}}{g^2} + c_0 \hbar + c_1 \hbar^2 g^2
+ \cdots
\ee
where we have temporarily reintroduced the dependence on
$\hbar$. Note that the n-loop $\zeta_p$ will necessitate the
evaluation of the (n+1)-loop renormalization group coefficient
functions $\beta(g^2),\gamma_2(g^2)$ and $\delta(g^2)$. If we
put $\alpha = 0$, we not only eliminate an independent parameter
but the vacuum energy divergences become
multiplicatively renormalizable :
\be
\zeta(g^2) + \delta \zeta(g^2,\epsilon) =
Z_{\zeta}(g^2,\epsilon)\zeta(g^2) 
\ee
Since $\zeta$ is now a unique function of $g^2$ which runs
according to the RGE, the energy functional W(J) obeys the
homogeneous RGE :
\be
\left( \mu \frac{\partial}{\partial \mu} + \beta(g^2)
\frac{\partial}{\partial g^2} - \gamma_2(g^2) \int d^4 x J
\frac{\delta}{\delta J}\right) W(J) = 0
\ee
Therefore the composite operator
\be
\frac{1}{2} Z_2 A^2_{\mu} -Z_{\zeta} \zeta J
\ee
has a finite and multiplicatively renormalizable expectation
value $ \Delta_R = \frac{\delta W}{\delta J}$ and two-point
function. For $J = 0, \Delta_R = 0$ on the perturbative sheet
while $\Delta_R = Z_2 \Delta_{np}$ on the non-perturbative
sheet. The effective action for $\Delta_R$ is defined by
\be
\Gamma(\Delta_R) = W(J) - \int d^4 x J \Delta_R
\ee
and obeys the RGE
\be
\left( \mu \frac{\partial}{\partial \mu} + \beta(g^2)
\frac{\partial}{\partial g^2} + \gamma_2 (g^2)\int d^4 x
\Delta_R \frac{\delta}{\delta \Delta_R}\right) \Gamma(\Delta_R)
= 0
\ee

To calculate $\Gamma(\Delta_R)$ one can proceed in a
straightforward way by calculating W(J) and doing the inversion.
This is rather cumbersome though, especially for space-time
dependent J. A much more efficient method which displays
explicitely the energy interpretation of $\Gamma(\Delta_R)$ uses
a Hubbard-Stratonovich transformation
\be
1 = \int [d\sigma]\exp - \frac{1}{2 Z_{\zeta}\zeta} \int d^D x
\left[ \frac{\sigma}{g} + \frac{1}{2}\mu^{\epsilon/2} Z_2 A^2_{\mu} -
\mu^{-\epsilon/2} Z_{\zeta} \zeta J \right]^2
\ee
to eliminate the $\frac{1}{2} Z_2 J A^2_{\mu}$ and $Z_{\zeta} \zeta J^2$
terms from the Lagrangian. Our energy functional can now be
written as a pathintegral over $A_{\mu}$ and $\sigma$ fields
\be
e^{-W(J)} = \int [dA_{\mu}][d\sigma] \exp - \int \left[ {\cal
L}(A_{\mu},\sigma) - \frac{\sigma J}{g}\right] d^D x
\ee
where the $\sigma$-field Lagrangian is given by
\bea
{\cal L}(\sigma,A_{\mu}) 
& = & \frac{1}{4} (F^a_{\mu \nu})^2 + {\cal L}_{g.f.} + {\cal
L}_{c.t.}\nonumber \\
& + & \frac{\sigma^2}{2g^2 Z_{\zeta}\zeta} + \frac{1}{2} \mu^{\epsilon/2}
\frac{Z_2}{g^2 Z_{\zeta}\zeta} g \sigma A^a_{\mu}
A^a_{\mu} + \frac{1}{8} \mu^{\epsilon}
\frac{Z^2_2}{Z_{\zeta}\zeta} (A^a_{\mu} A^a_{\mu})^2
\eea
In our new expression for W(J), J appears now as a linear source
term for the $\sigma$ field so that $\langle \sigma \rangle = -
g\Delta_R$. The inversion and Legendre transform are therefore
unnecessary and we simply have
\be
\Gamma(\Delta_R) = \Gamma_{1PI} (\sigma = - g \Delta_R)
\ee
which can be calculated in perturbation theory using the
background field formalism.

We have obtained a new multiplicatively renormalizable
Lagrangian ${\cal L}(\sigma, A_{\mu})$ which is to all orders in
perturbation theory equivalent to the original Yang-Mills
Lagrangian. If one perturbs around $\sigma = 0$, one recovers
the original perturbation series with is well known problems
such a infrared renormalons. If one expands around $\sigma \neq
0$, one has an effective gluon mass which incorporates
non-perturbative effects signalled by the infrared renormalons.

To see whether the groundstate favours $\sigma \neq 0$, we have
calculated the effective potential for $\sigma$ up to two loops.
To calculate $\zeta(g^2)$ up to two loops, we had to calculate
the R.G. functions up to three loops. The calculations where
done in the Landau gauge in the $\overline{MS}$ scheme in $D = 4
- \epsilon$ using the tensor correction method [10] which is a
new method for efficient calculation of multiloop Feynman
diagrams. We calculated W(J) up to three loops and found that it
could be renormalised with the counterterm $- \delta \zeta
J^2/2$ where
\bea
\delta \zeta & = & \frac{(N^2_c - 1)}{16\pi^2} \left[ - \frac{3}{\epsilon}
+ \left( \frac{g^2 N_c}{16\pi^2}\right) \left( \frac{35}{2}
\frac{1}{\epsilon^2} -
\frac{139}{6} \frac{1}{\epsilon} \right) \right. \nonumber \\
& + & \left. \left( \frac{g^2 N_c}{16 \pi^2}\right)^2 \left( - \frac
{665}{6} \frac{1}{\epsilon^3} + \frac{6629}{36}
\frac{1}{\epsilon^2} - \left( \frac{71551}{432} + \frac{231}{16}
\zeta (3)\right) \frac{1}{\epsilon} \right) \right]
\eea
For mass renormalisation, we found
\bea
Z_2 & = & 1 - \left( \frac{g^2 N_c}{16 \pi^2}\right) \frac{3}{2\epsilon} +
\left( \frac{g^2 N_c}{16\pi^2}\right)^2 \left(
\frac{53}{8} \frac{1}{\epsilon^2} - \frac{95}{48} \frac{1}{\epsilon}\right) \nonumber
\\ 
& + & \left( \frac{g^2 N_c}{16\pi^2}\right)^3 \left( -
\frac{5141}{144} \frac{1}{\epsilon^3} + \frac{20717}{864}
\frac{1}{\epsilon^2} - \left( \frac{11713}{2592} +
\frac{3\zeta(3)}{16}\right) \frac{1}{\epsilon}\right)
\eea
and anomalous dimension :
\be
\gamma_2 (g^2) = \left( \frac{g^2 N_c}{16\pi^2}\right) \frac{35}{6} + \left(
\frac{g^2 N_c}{16\pi^2}\right)^2 \frac{449}{24} + \left(
\frac{g^2 N_c}{16\pi^2}\right)^3 \left( \frac{94363}{864} -
\frac{9}{16} \zeta(3)\right)
\ee
For the renormalisation group function of the vacuum energy, we
obtained using (8), (21) and (23) :
\be
\delta(g^2) = \frac{(N^2_c -1)}{16\pi^2} \left[ -3 - \left(
\frac{g^2 N_c}{16\pi^2}\right) \frac{139}{3} - \left( \frac{g^2
N_c}{16\pi^2}\right)^2 \left( \frac{71551}{144} + \frac{693}{16}
\zeta (3)\right) \right]
\ee
Finally we solved (10) with a Laurent expansion in $g^2$ and
found up to two loops :
\be
\zeta (g^2) = \frac{(N_c^2 - 1)}{16\pi^2} \left[ \left(
\frac{16\pi^2}{g^2 N_c}\right) \frac{9}{13} + \frac{161}{52} + 
\left( \frac{g^2 N_c}{16\pi^2}\right) \left( \frac{567343 +
82539 \zeta (3)}{35568} \right) \right]
\ee
From (21) and (25) one can calculate $Z_{\zeta}$, so we now have
all ingredients to calculate ${\cal L}(\sigma,A_{\mu})$ to two
loop order.

We can read off the effective gluon mass in lowest order from
(19) and (25) : 
\be
m^2_{eff} = g\sigma \frac{13}{9} \frac{N_c}{N^2_c - 1}
\ee
We define $\sigma^{\prime} = \frac{13}{9} \frac{N_c}{N^2_c -
1} \sigma$ so that the background field method at one loop gives
free gluons propagating with an effective mass $m^2_{eff} = g
\sigma^{\prime}$. Since at one loop, $Z_{\zeta} = 1 -
\frac{13}{3} \left( \frac{g^2 N_c}{16\pi^2}\right)
\frac{1}{\epsilon}$ and using the one loop value of $\zeta(g^2)$, we have
\bea
V_1(\sigma^{\prime}) & = & \frac{9}{13} \frac{(N^2_c - 1)}{N_c}
\frac{\sigma^{\prime 2}}{2} \left[ 1 + \frac{13}{3} \left( \frac{g^2
N_c}{16\pi^2}\right) \frac{1}{\epsilon} - \frac{13}{9} \frac{161}{52}
\left( \frac{g^2 N_c}{16\pi^2}\right)\right] \nonumber \\
& + & \frac{1}{2} Tr \ln (- \Box + g^2 \sigma^{\prime 2})
\eea
where the trace goes over color and Lorentz indices. Because there
are $N^2_c - 1$ gluons with 3 massive polarizations in the
Landau gauge, we find in het $\overline{MS}$ scheme :
\be
\frac{1}{2} Tr \ln (- \Box + g^2 \sigma^{\prime 2}) =
\frac{3(N^2_c - 1)}{64\pi^2} g^2 \sigma^{\prime 2} \left[ -
\frac{2}{\epsilon} - \frac{5}{6} + \ln
\frac{g\sigma^{\prime}}{\overline{\mu}^2} \right]
\ee
The divergences cancel and we obtain a finite one loop effective
potential :
\be
V_1(\sigma^{\prime}) = \frac{9}{13} \frac{(N^2_c - 1)}{N_c}
\frac{\sigma^{\prime 2}}{2} + \frac{3}{4} (N^2_c - 1)
\frac{(g\sigma^{\prime})^2} {16\pi^2} \left[ - \frac{5}{6} -
\frac{161}{78} + \ln
\frac{g\sigma^{\prime}}{\overline{\mu}^2}\right] 
\ee
The two loop correction has been calculated in [10] and reads :
\bea
\Delta V_2(\sigma^{\prime}) & = & (N^2_c - 1)
\frac{(g\sigma^{\prime})^2}{16\pi^2} \left( \frac{g^2
N_c}{16\pi^2}\right) \left[ \frac{21}{4} \ln
\frac{g\sigma^{\prime}}{\overline {\mu}^2} - \frac{9}{16} \left(
\ln \frac{g\sigma^{\prime}}{\overline{\mu}^2}\right)^2 \right.\nonumber
\\
& - & \left. \frac{49359}{3952} + \frac{891}{32} s_2 -
\frac{\zeta(2)}{16} - \frac{9171}{7904} \zeta (3) \right]
\eea
where $s_2 = \frac{4}{9\sqrt{3}} C\ell_2 (\pi/3) \simeq
0.2604341 \cdots$.

At one loop as well as at two loops, the perturbative vacuum
$\sigma^{\prime} = 0$ is a local maximum and a lower minimum is
obtained for $\sigma^{\prime} \neq 0$. We can use the RGE to sum
leading logarithms and put $\overline{\mu}^2 = g
\sigma^{\prime}$. Introducing the expansion parameter $y =
\frac{g^2 N_c}{16\pi^2}$ we find a global minimum for
$V(\sigma^{\prime})$ at one loop for $y_1 = 0.19251$ and at two
loops for $y_2 = 0.14466$, independent of $N_c$. The
corresponding coupling constants are reasonably small : for $N_c
= 3, \alpha \sim0.8$ (1 loop) or $\alpha \sim0.6$ (2 loops).
Through dimensional transmutation we obtain non-vanishing
effective gluon masses. At one loop we find $m_1 = (g\sigma)_1
\approx 2.05 \Lambda_{\overline{MS}} \sim 485 \mbox{MeV}$ for
$\overline{\Lambda}_{MS} = 237 \mbox{MeV}$. At two loops, we find using
the one loop $\beta$-function $m_{21} \approx 2.59
\Lambda_{\overline{MS}} \sim 614 \mbox{MeV}$ and using the two loop
$\beta$-function, $m_{22} \approx 1.96 \Lambda_{\overline{MS}}
\sim 464 \mbox{MeV}$. For the non-perturbative vacuum energy density
and for $N_c = 3$ we find $\epsilon^1_{vac} \approx - 0.335
\Lambda^4_{\overline {MS}}$ at one loop while at two loops we
find, $\epsilon^{21}_{vac} \approx - 1,7
\Lambda^4_{\overline{MS}}$ and $\epsilon^{22}_{vac} \approx -
0.567 \Lambda^4_{\overline{MS}}$.
Finally we can calculate the gluon condensate $\langle
\frac{\alpha_s}{\pi} F^2\rangle$ by making use of the trace
anomaly :
\be
\Theta_{\mu \mu} = \frac{\beta (g)}{2g} \left( F^a_{\lambda
\sigma}\right)^2 
\ee
From the anomaly we deduce for $N_c = 3$ that the gluon
condensate is related to the vacuum energy density as :
\be
\langle \frac{\alpha}{\pi} F^2 \rangle = - \frac{32}{11}
\epsilon_{vac} 
\ee
Using our numerical results for $\epsilon_{vac}$, in one and
two loops (with one and two loop $\beta$-functions) we find for
the gluon condensate :
\bea
\langle \frac{\alpha}{\pi} F^2 \rangle_1 & = & 0.0031 {\mbox GeV^4}
\nonumber \\
\langle \frac{\alpha}{\pi} F^2 \rangle_{21} & = & 0.0156 {\mbox GeV^4}
\nonumber \\
\langle \frac{\alpha}{\pi} F^2 \rangle_{22} & = & 0.0052 {\mbox GeV^4}
\eea
Since $\epsilon_{vac}$ in (32) is really the energy difference
between the non-perturbative and the perturbative groundstate,
our definition of the gluon condensate is in fact $ \langle
\frac{\alpha}{\pi} F^2\rangle = \langle \frac{\alpha}{\pi} F^2
\rangle_{np} - \langle \frac{\alpha}{\pi} F^2 \rangle_p$ where
the suffices p and np means taking the J = 0 limit on the
perturbative and non-perturbative sheet respectively.

In this letter we have introduced a consistent definition for
the non-perturbative value of the local composite operator $
A^2_{\mu}$ and given evidence through two loop calculations of a
multiplicatively renormalisable effective potential that the
non-perturbative vacuum favours a non-zero value for this
condensate. Our calculations can only be seen as qualitative
indications that non-perturbative values for $A^2_{\mu}$ can
lower the energy. Other important non-perturbative effects such
as instantons have been left out in the calculation of the
effective potential. Since the operator $\tilde{A}^2_{\mu}$
from which we start in equation (1) is gauge invariant, our
results are gauge invariant. For a non zero condensate $\sigma
\neq 0$, perturbative unitarity in the gluon sector is broken.
It is known [12], that gauge invariance and perturbative
unitarity should not always go together. However, confinement
could solve this and secure non-perturbative unitarity in the
zero color sector.


\newpage
\begin{thebibliography}{99}
\bibitem{Gubarev}
F.V. Gubarev, L. Stodolsky and V.I. Zakharov, Phys.Rev.Lett.
{\bf 86} (2001) 2220
\bibitem{Gubarev}
F.V. Gubarev and V.I. Zakharov, Phys.Lett. {\bf B501} (2001) 28 \\
K.G. Chetyrkin, S. Narison and V.I. Zakharov, Nucl.Phys. {\bf
B550} (1999) 353
\bibitem{David}
F. David, Nucl.Phys. {\bf B263} (1986) 637
\bibitem{Mueller}
A.H. Mueller, Nucl.Phys. {\bf B250} (1985) 327
\bibitem{Banks}
T. Banks and S. Raby, Phys.Rev. {\bf D14} (1976) 2182
\bibitem{Veltman}
M. Veltman, Nucl.Phys. {\bf B21} (1970) 288 \\
D.G. Boulware, Ann.Phys. {\bf 56} (1970) 140
\bibitem{van Dam}
H. van Dam and M. Veltman, Nucl.Phys. {\bf B22} (1971) 397 \\
V.I. Zakharov, JETP Lett. {\bf 12} (1970) 312
\bibitem{Gross}
D. Gross and A. Neveu, Phys.Rev. {\bf D10} (1974) 3235
\bibitem{Verschelde}
H. Verschelde, Phys.Lett. {\bf B351} (1995) 242 \\
H. Verschelde et al., Z.Phys. {\bf C76} (1997) 161
\bibitem{Knecht}
K. Knecht, Ph.D thesis (Univ. of Gent)
\bibitem{Zwanziger}
D. Zwanziger, Nucl.Phys. {\bf B412} (1994) 657
\bibitem{delbourgo}
R. Delbourgo and G. Thompson, Phys.Rev.Lett. {\bf 58} (1987) 2001\\
R. Delbourgo, S. Twisk and G. Thompson, Int.J.Mod.Phys. {\bf A3} (1988) 435
\end{thebibliography}
\end{document}